\documentclass[dvips]{acta}
\usepackage{supertabular,lscape,epsfig}
\usepackage{amssymb}
\usepackage{amsmath}

\SetPages{0}{0}
\SetVol{66}{2016}

\begin{document}

\def\sv{\scriptscriptstyle\rm V}
\def\sb{\scriptscriptstyle\rm B}
\def\su{\scriptscriptstyle\rm U}

\begin{Titlepage}
\Title{The appearance of non-spherical systems. 
Application to LMXB}

\Author{A.~~ R\'o\.za\'nska$^1$, B.~~Be{\l }dycki$^1$, J.~~M~a~d~e~j$^2$,
 T.P.~~Adhikari$^1$ and B. You$^1$}
{$^1$ N. Copernicus Astronomical Center, Bartycka 18, 00-716 Warsaw, Poland,      \\
e-mail: agata@camk.edu.pl    \\
$^2$ Warsaw University Observatory, Al. Ujazdowskie 4, 00-478 Warszawa, Poland \\
e-mail: jm@astrouw.edu.pl
}

\Received{Month Day, Year}
\end{Titlepage}

\Abstract{We study the appearance of the neutron star - accretion disk system as seen by 
a distant observer in the UV/X-ray domain. The observed intensity spectra are computed assuming
 non-spherical geometry of the whole  system, in which outgoing spectrum is not represented by the flux spectrum, the latter being valid for spherically symmetric objects. Intensity spectra of our model  display double bumps in UV/X-ray energy domains. 
 Such structure is caused by the fact that the 
the source is not spherically symmetric, and the proper integration of intensity over emitted 
area is needed to reproduce observed spectral shape. Relative normalization of double bump is self consistently computed by our model. X-ray spectra of such a type were often observed in LMXB with accretion disk, ultra luminous X-ray sources, and accreting black hole systems with hot inner compact corona. Our model naturally explains high energy broadening of the disk spectrum observed in some binaries. We attempted to fit our model to X-ray data of XTE~J1709-267 from {\it XMM-Newton}. Unfortunately, the double  intensity bump predicted by our model for LMXB is located in soft X-ray domain, uncovered by existing data for this source.
}

{accretion,accretion disks -- stars:neutron -- X-ray:binaries}

\section{Introduction}

Low mass X-ray binaries (LMXB) are binary systems, where the primary star is 
 a  compact object, being the black hole
or the neutron star.  The secondary is a late type  main sequence star, 
classified as K or M of low mass  slightly less than the solar mass, $M_{\odot}$. 
In many of those sources  accretion of matter onto the compact object occurs 
if the system is tight enough.
Accretion can proceed under different scenarios. In principle, when falling 
matter has non-zero initial angular momentum, then the accretion disk may form down 
to the innermost  marginally stable orbit (ISCO).  

Since the time when the standard accretion disk model was defined by Shakura \& Sunyaev (1973), 
theoreticians developed the idea taking into account relativistic corrections 
(Novikov \& Thorne 1973) and radial advection (Abramowicz et al. 1988).
According to all those models, accretion disks in LMXB are hot in their central 
part, having the gas temperature of the order of $10^{6-7}$ K.
For the purpose of this paper, relativistic corrections and radial advection were neglected, 
since we model basic disk emission.
If the compact object in LMXB is a neutron star, then we should expect 
that the neutron star is hot, with $T_{\rm eff,NS}$ up to  a few $10^{6-7.5}$ K.  
Such system is bright in X-ray domain, and the derivation of the total shape 
of its intensity spectrum is the principal aim of this paper. 

In this research project we computed the observed intensity of the whole LMXB 
system containing a neutron star with the accretion disk. 
The secondary is not visible in X-rays and that star is  only important to estimate 
the size of the Roche lobe, and therefore, the outer disk radius. 
 Since the emitting source such as LMXB is not spherically symmetric then we have 
to reject standard formula that the observed luminosity is just proportional to the flux 
(in ergs per second per one cm$^{2}$) emitted from the unit surface of a spherical star
(Mihalas 1978). In case of the disk-like system we have to compute
the observed intensity starting from the basic formula i.e. Eq.~[1-27].
in Mihalas (1978). In this paper we derived such formula 
appropriate to the LMXB with neutron star, but the same calculation can be done 
for any non-spherical system.

We calculated the observed intensity assuming that both 
the neutron star and the accretion disk radiate as black bodies. The radial effects of mutual 
attenuation were fully taken into account. 
Our model spectra seen at different viewing angles  display double bumps in 
the UV/X-ray energy domain. The relative strength of bumps depends on the neutron star and the 
inner accretion disk effective temperatures.  Furthermore, the high energy bump originating 
from partially attenuated neutron star depends on the viewing angle, which is not the case of 
the neutron star alone. 
Such a type of continuum spectra are very often observed in LMXB with accretion 
disks,  ultra luminous X-ray sources (ULXs), and black hole accretion disks with hot corona. 
Our model should be used for any non-spherical systems containing the accretion disk and the inner emitting source of different temperature, 
which can be for example, the hot compact corona (Fabian et al. 2015).

In Sec.~2
we present the source geometry and derive 
the observed intensity of the whole system as a function of the viewing angle. 
The resulting appearance of non-spherical systems with parameters typical for 
LMXB is drawn in Sec.~3.
The first fit of our models to the X-ray data of XTE 
J1709-267 is shown in Sec.~4. 
We discuss and conclude  our work in Sec.~5.

\section{Model geometry}
\label{sec:mod}

\begin{figure*}
\begin{center}
\includegraphics[scale=0.45]{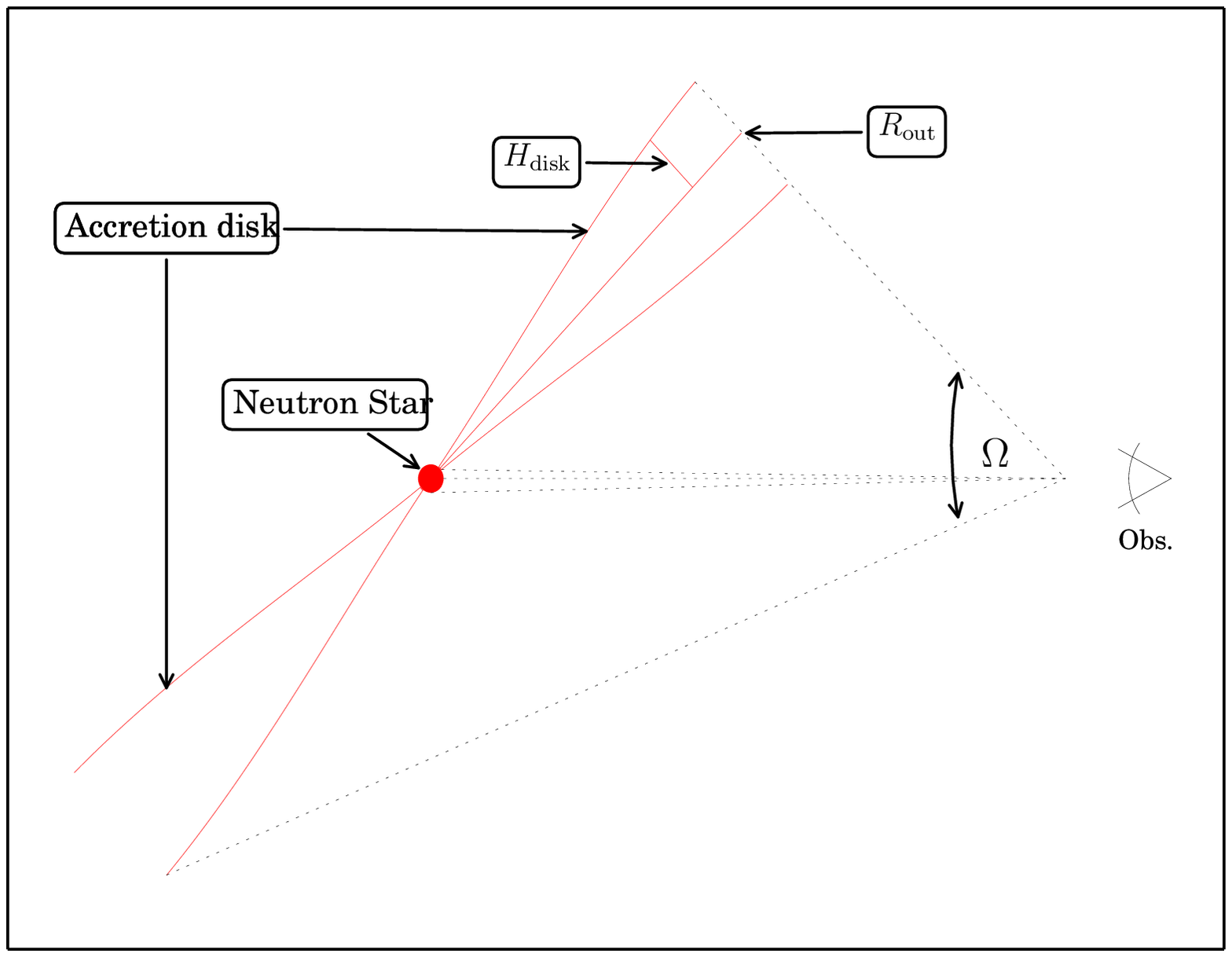}
\end{center}
\caption{Sketch of the LMXB with the neutron star. All 
geometrical proportions are set according to our assumptions, see 
Sec.~3 for exact values of parameters. }
\label{fig:one}
\end{figure*}

\subsection{Glossary of terms}
We consider below the following variables: specific intensity, flux of radiation and the 
observed intensity, as standard variables used in model atmosphere calculations. Nevertheless, 
we draw attention of the reader to differences between exact meaning of those variables. 
\begin{enumerate}

\item Specific intensity $ I_{\nu} $ as defined by Mihalas (1978) as the energy of radiation 
which flows through one cm$^2$ on the surface of an emitter into a direction, which generally 
is inclined to the star's surface. Intensity  $ I_{\nu} $ is an intrinsic property of the source.

\item Energy dependent flux of radiation in plane parallel geometry:
$F_{\nu}=\oint I_{\nu} d\omega$  is the average of the specific intensity $ I_{\nu} $ 
weighted by cos$\,\theta$ (zenithal angle),
and represents the energy flux flowing through one cm$^{2}$ on the surface of the emitter.
Integration is performed over the full solid angle $4\pi$.
In other words this flux represents intrinsic property of the source of radiation. 

\item  At the same time, infinitesimal energy $d \mathcal{F_{\nu}}$ can be 
measured by a distant observer, because specific intensity does not change in the 
empty space along the ray path. That energy is defined as:
\begin{equation} 
d \mathcal{F_{\nu}} = I_{\nu} d\omega,
\label{row:1}
\end{equation}
where $ d\omega $ is a solid angle in steradian [sr]. 
This formula is applicable for the flat space, and we assume it in this paper. 
In case when the emitter is located close to the black hole, both general and special
 relativistic corrections should be 
taken into account as shown by Fabian et al. (1989, Eq.~A4).

Integrating Eq.~1 over the solid angle subtended by the source, 
we obtain energy dependent intensity: $ \mathcal{F_{\nu}}$
per unit time measured by one cm$^{2}$ of the detector.  This quantity has the same unit as
 energy dependent flux and diminishes with the increasing distance to the observer. But it 
is not an intrinsic property of the source, and we show below how it depends on the source 
geometry. 

\end{enumerate}

\subsection{The observed intensity}
\label{sec:lum}
\begin{figure*}
\begin{tabular}{ll}
\hspace{-1cm} \includegraphics[scale=0.4]{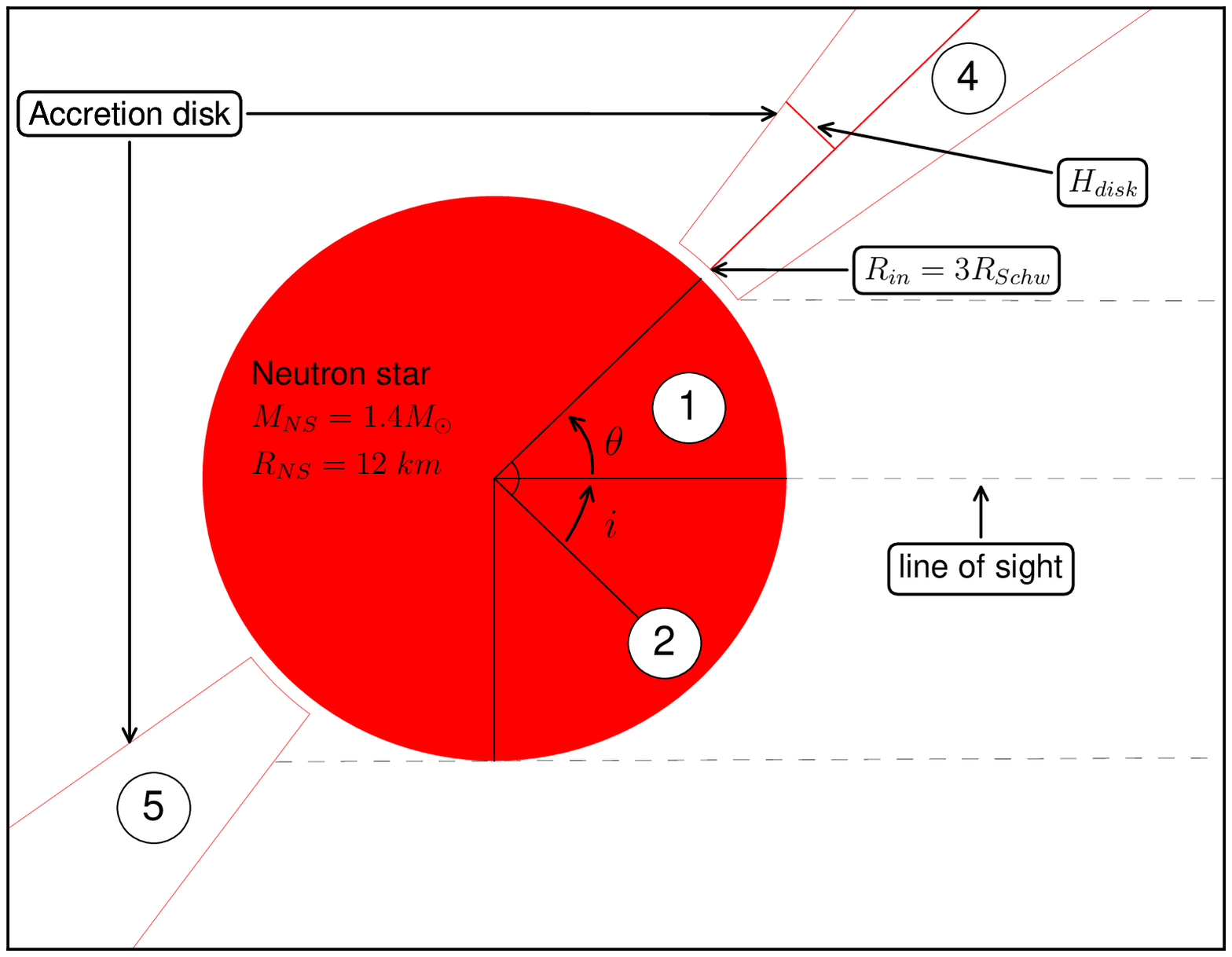} &
\hspace{-1cm} \includegraphics[scale=0.4]{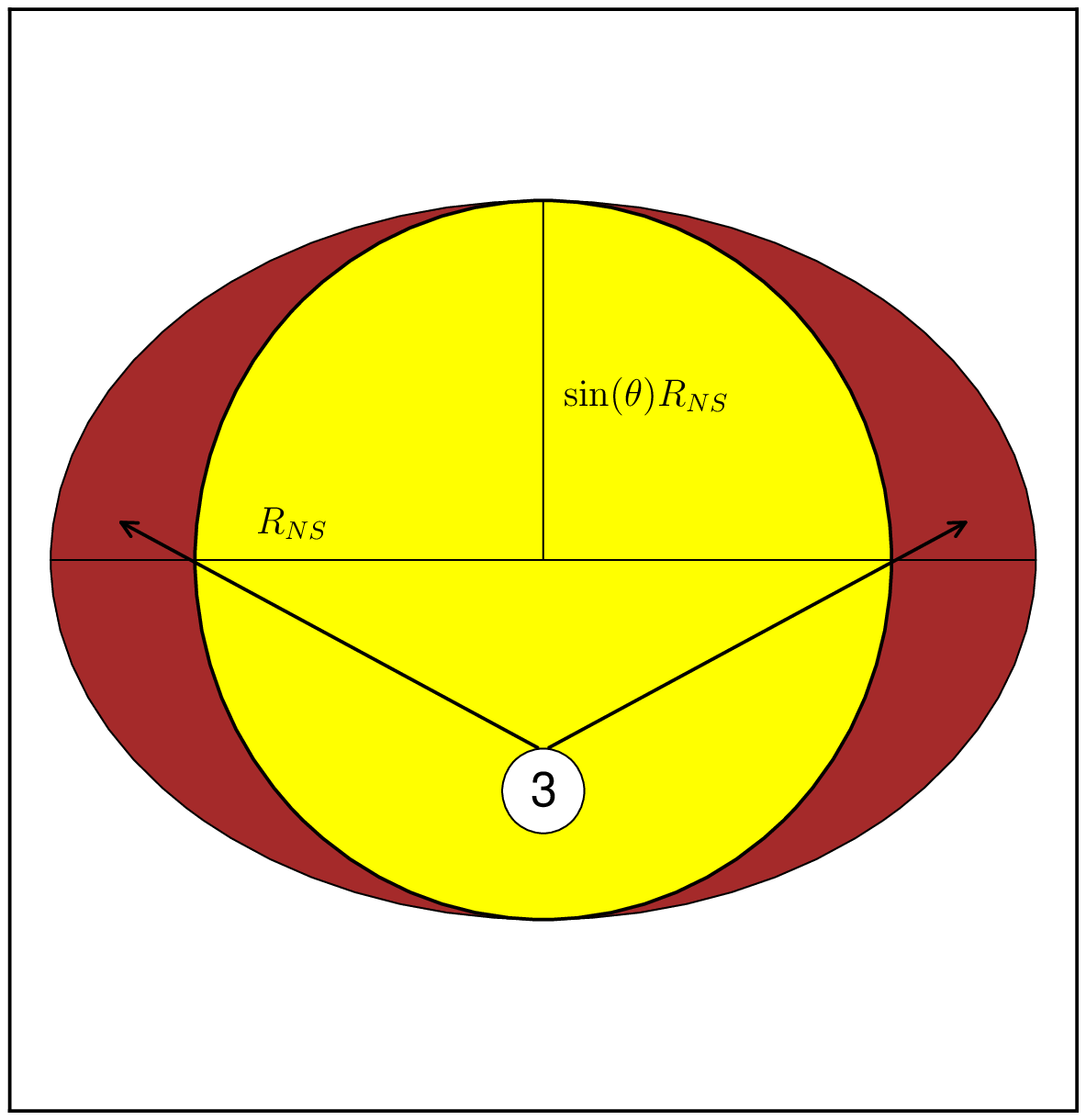} \\
\end{tabular}
\caption{Close look on the LMXB with the neutron star as a compact object. 
Distant observer sees whole system at aspect 
angle $i=90-\theta$. Observed 
intensity takes into account all mutual attenuation effects. Numbers mark different emitting parts from which total observed intensity is computed with the use of Eq.~6.}
\label{fig:stch}
\end{figure*}

Assuming that the emitter is  spherically symmetric neutron star surface, we 
can define its observed monochromatic intensity  following standard formula derived by 
by Mihalas (1978) (Eq.~1-27):
\begin{equation}
\mathcal{F_{\nu, \rm NS}}=\int_{\Omega} I_{\nu} d\omega
\end{equation}
where $I_{\nu}$ denotes specific intensity and 
$\Omega$ is the solid angle subtended by the area as seen by an observer. 
The integration undergoes over solid angle which is small enough
to substitute  $d \omega = dS/ D^2$, where dS is the 
 surface of the star defined as an annulus $dS=2 \pi \, r \, dr$ 
normal to the line of sight, and $D$ is the distance to the observer. 
For spherically symmetric neutron star with the radius $R_{\rm NS}$ 
we get: $r=R_{\rm NS} \sin {\theta}$, where $\theta$ is the angle between direction of the light beam and the normal to the surface. Since $\mu = \cos{\theta}$,
the solid angle becomes
$d \omega = 2 \pi (R_{\rm NS}/D)^2 \,\mu \,d\mu$, and finally we obtain:  
\begin{equation}
\mathcal{F_{\nu, \rm NS}}= 2 \pi  \left({R_{\rm NS} \over D}\right)^2 \, 
    \int_0^1 I_{\nu} \mu d\mu = 
  \left({R_{\rm NS} \over D} \right)^2 F_{ \nu}.
\label{eq:nstar}
\end{equation}  
We note here, that the observed intensity per detector area is proportional  to the  flux 
emitted from 1~cm$^{2}$  of the star's  surface,
 only due to the spherical shape of the emitting region. 

In case of axisymmetric standard geometrically thin accretion disk 
(for example: Shakura \& Sunyaev 1973, hereafter SS73), 
the same procedure gives:
\begin{equation}
\mathcal{F_{\nu, \rm AD}} =\int_{\Omega} I_{\nu} d\omega = 2 \pi {{sin{\theta} } \over {D^2}}
\int_{R_{\rm in}}^{R_{\rm out}} I_{\nu} R dR,  
\label{eq:disk}
\end{equation}
where the monochromatic intensity, $I_{\nu}$ emitted in the specific direction is integrated 
over the whole disk surface from the inner to outer disk radii. 
This algebraic calculation uses the fact that the projected disk area 
equals $dS= 2 \pi\, \sin{\theta}\, R\, dR$, 
and the integration undergoes over $d\omega= 2 \pi/ D^2 \, \sin{\theta} \, R\, dR $. 

In the case of emission from the whole system i.e. neutron star with the accretion disk
around it, we have the contribution from different emitting parts, as shown in Fig.~1. 
In such non-spherical source, the total observed monochromatic intensity is given by: 
\begin{equation}
\mathcal{F_{\nu, \rm All}}= \mathcal{F_{\nu, \rm NS}^{\rm 1}}+\mathcal{F_{\nu, \rm NS}^{\rm 2}}+
\mathcal{F_{\nu, \rm NS}^{\rm 3}} + \mathcal{F_{\nu, \rm AD}^{\rm 4}} + \mathcal{F_{\nu, \rm AD}^{\rm 5}}.
\end{equation} 
The numbers correspond to the emitting regions clearly shown in Fig.~2.
 The first term of above equation, $\mathcal{F_{\nu, \rm NS}^{\rm 1}}$, represents the 
half of neutron star emission given by Eq.~3, the next two terms describe  
emission from the half of a smaller circle, $\mathcal{F_{\nu, \rm NS}^{\rm 2}}$, 
plus the emission from {\it ears}, $\mathcal{F_{\nu, \rm NS}^{\rm 3}}$, which are formed 
on the both sides of smaller circle due to the disk attenuation. 
The first part of the  disk emission, $ \mathcal{F_{\nu, \rm AD}^{\rm 4}}$ is 
given by the  half of $\mathcal{F_{\nu, \rm AD}}$ resulting by Eq.~4, while the second  term, $\mathcal{F_{\nu, \rm AD}^{\rm 5}}$,
is the same  $\mathcal{F_{\nu, \rm AD}}$ integrated over $R_{\rm boost}= R_{\rm NS} \, \sin{\theta}$ up
to $R_{\rm out}$, due to neutron star attenuation. 

The final observed energy dependent intensity directed
to the observer is computed as: 
\begin{eqnarray}
\mathcal{F_{\nu, \rm All}} & = & \pi \left({R_{\rm NS} \over D}\right)^2 \, \left[ 
\int_0^1 I_{\nu} \mu d\mu + \int_{\cos{\theta}}^1 I_{\nu} \mu d\mu \right]  \nonumber \\
&+   &  { 2 \over D^2}  \left[ \int_0^{R_{\rm NS}} I_{\nu} \sin{\theta} \, \sqrt{R_{\rm NS}^2-x^2} \, dx  \right.   \nonumber \\ 
& -&  \left. \int_0^{R_{\rm boost}} I_{\nu}  \sqrt{R_{\rm boost}^2-x^2}\, dx  \right]  \nonumber \\
&+& \pi {{\sin{\theta} } \over {D^2}}  \left( \int_{R_{\rm in}}^{R_{\rm out}} I_{\nu} R dR 
+  \int_{R_{\rm boost}}^{R_{\rm out}} I_{\nu} R dR \right),
\label{eq:all}
\end{eqnarray}
where $x$ denotes the variable of integration over the star surface projected on the sky. 
Note, that the angle ${\theta}$ is related to the disk viewing angle as $i=90^{\circ}- \theta$ (see Fig.~2 for illustration). In contrast to neutron star atmosphere, 
the observed intensity for non-spherical sources is not proportional to the locally emitted flux 
(Eq.~4, and~6).

We assume, that the specific intensity of the neutron star radiation is isotropic 
and equals  the black body intensity at the given effective 
temperature $I_{\nu}= B_{\nu} (T_{\rm eff,NS})$. Furthermore, we assume that the  specific
intensity emitted at different disk radii is isotropic and equals to the local 
Planck function  $I_{\nu}=B_{\nu}(T_{\rm eff}(R))$, with effective temperature  at  
the disk radius given by standard SS73 formula:
$\sigma T_{\rm eff}^4 = {3 GM \dot M / 8 \pi  R^3} 
( 1- ({R_{in} / R } )^{1/2} )$,   where $G$ is the gravitational constant, $M$ - mass of the central object and $\dot M$ - disk accretion rate. 
We note here, that our model is useful  for systems where the angle dependent 
specific intensity is given by results of the radiative transfer calculations
(Madej 1991; Hubeny et al. 2001; Davis et al. 2005; R\'o\.za\'nska et al. 2011).

\section{LMXB at different viewing angles}
\label{sec:results}

In this section we demonstrate that the broad-band emitted spectrum from non-spherical 
system depends on the viewing angle in the way that both components: disk emission and 
neutron star emission change with angle. Since we intend to compare our models to the observed 
X-ray spectra of the LMXB source, 
we constructed the grid of models for arbitrarily assumed parameters. The non-rotating neutron 
star  has a canonical mass 1.4~M$_{\odot}$,  radius 
12~km, and 11 various effective temperatures, ranging from $2\times 10^6$~K up 
to $4\times 10^7$~K.  The disk local emission was computed assuming 11 accretion rates
  from $8 \times 10^{-4}$ up to $8 \times 10^{-1}$ in the unit of the  Eddington accretion
 rate, with accretion efficiency equal 0.08 (Schwarzschild metric). 

For each disk model, we calculated multi-black body spectrum from $R_{\rm in}$ to $R_{\rm out}$
 between the range 3-1000 $R_{\rm Schw}$, 
where the $R_{\rm Schw}=2GM_{\rm NS}/c^2$. The inner disk radius can change due to the: large 
value of magnetic field, strong boundary layer, and when the relativistic corrections are 
taken into account. We plan to do it in the future paper together with full ray tracing procedure.
 The outer disk radius depends on the secondary object, and resulting size of the Roche lobe. 
We have checked that for the typical mass of the secondary less than the solar mass, there is 
enough space for the disk of a radius up to 1000 $R_{\rm Schw}$ (Paczy\'nski 1971). Since 
the outer disk regions emit in optical band, it's value  does not influence our results. 
Therefore,  we keep the value of outer radius constant within this paper. 

\begin{figure}[h]
\begin{tabular}{ll}
\hspace{-0.5cm} \includegraphics[scale=0.4]{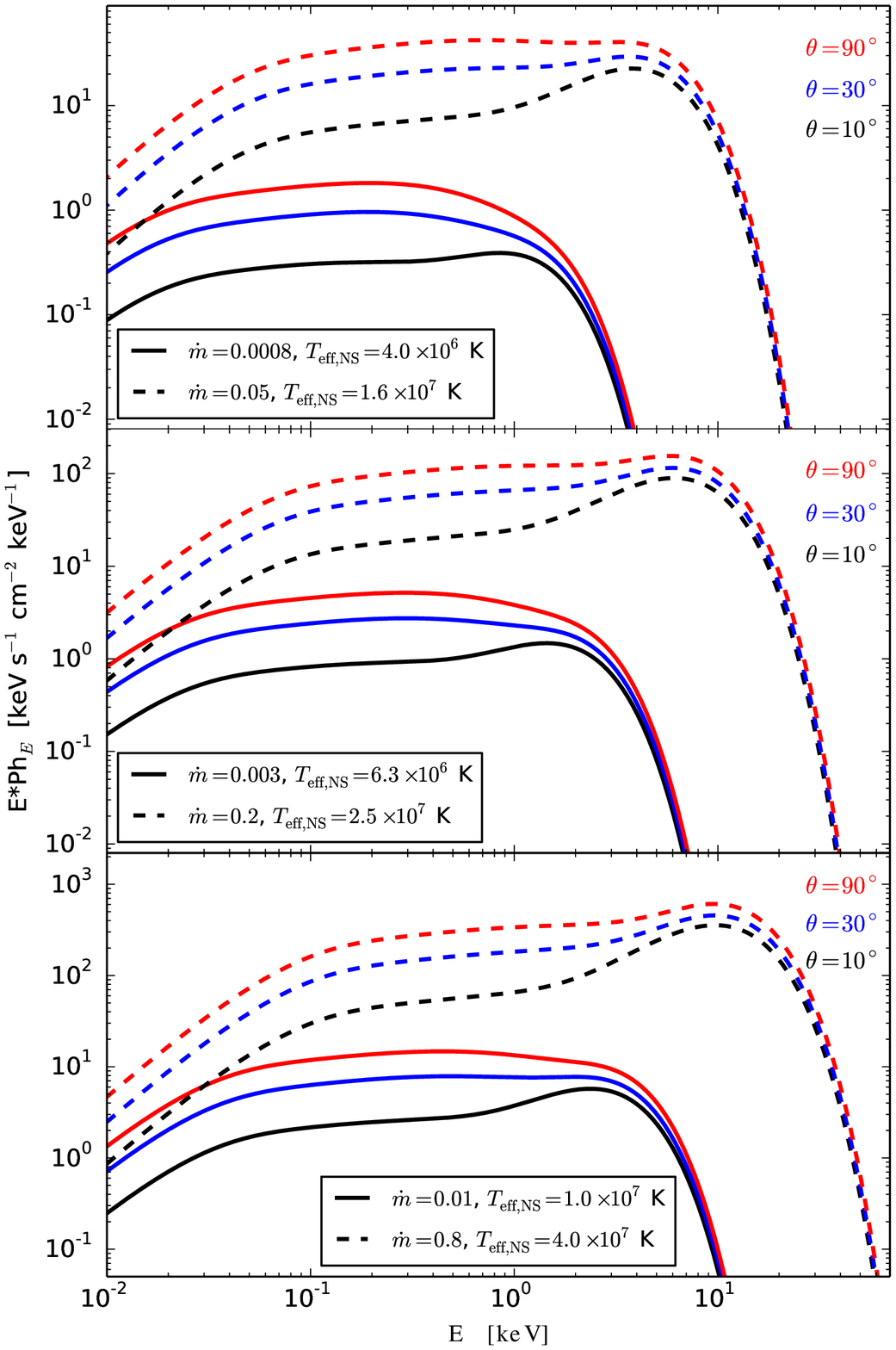} &
\hspace{-1cm} \includegraphics[scale=0.4]{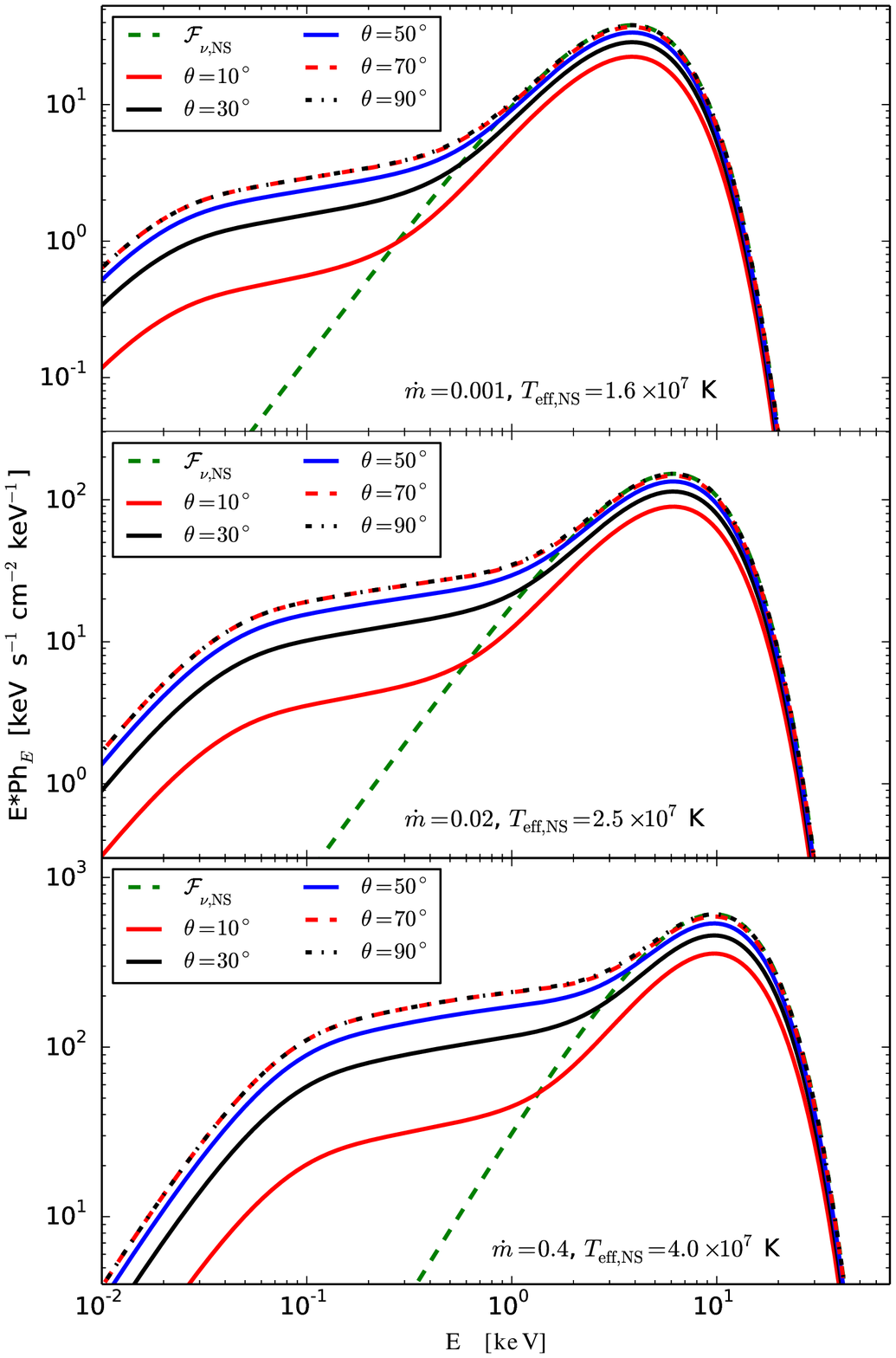} \\
\end{tabular}
 \caption{Observed intensity of non-spherical system in case of LMXB for different accretion 
rates, neutron star temperatures and inclinations. The dashed line at each 
right panel represents the emission from neutron star only given by the formula Eq.~3.}
 \label{fig:tot}
 \end{figure} 
 

The grid of $\theta$ angles spans from $10^{\circ}$ up to $90^{\circ}$. 
The lowest value of this angle corresponds almost to the ``edge on'' disk, while the highest 
value to ``face on'' disk. 
We use the lowest value of $\theta=10^{\circ}$, since for smaller viewing angles 
the disk geometrical thickness is large enough to cover neutron star completely. Furthermore, 
in SS73  model, the disk height increases with radius, and for the disk seen ``edge on'',  
we observe only the disk rim. We are aware of this problem, but to show the basic mutual 
attenuation effect we neglect the disk height. 
 
 The appearance of non-spherical system containing disk and neutron star for various parameters 
is presented in Fig~3. All the computed models are presented in units of the photon 
observed intensity which is the photon number multiplied by photon energy in keV, i.e.: 
keV~s$^{-1}$~cm$^{-2}$~keV$^{-1}$ (for direct comparison with observations).
 Presented spectra are broad, and the double bump structure in the observed intensity spectrum 
is clearly visible for some set of parameters. We associate low energy bump with the emission 
from an accretion disk, while high energy bump reflects the emission from hot neutron star. The
 observed intensity spectrum depends on the viewing angle in the whole energy range. 
 
 In Fig.~3 left panels we show observed spectra from the systems when neutron star emission 
dominates, i.e. when it's effective temperature is considerably larger than inner disk temperature.
 Furthermore, the emission from the neutron star alone, given by the formula Eq.~3, 
is presented by the green dashed line. It is clear that the level of neutron star emission depends
 on the viewing angle. 
Note, that our model was computed for the peculiar non-spherical system, but generally our 
approach should  be applied to any other set of emitting regions, for instance, for the disk with
 hot inner compact corona,  recently observed by {\it NuSTAR} X-ray telescope (Fabian et al. 2015).

\section{X-ray observations of XTE J1709-267}
\label{sec:obs}

To discuss the observational verification of our model we 
used the data on LMXB -- XTEJ1709-267 from {\it XMM-Newton} telescope, already published in  
Degenaar et al. (2013). The data 
were taken during 31 ks, and have observational ID $0700381401$. 
The observation consists of 11 exposures with 
all the available instruments of {\it XMM-Newton}. For this analysis, we 
selected  the data obtained with European Photon Imaging Camera (EPIC): 
MOS1, in the small window imaging mode. We used {\it XMM-Newton} Science 
Analysis Software (SAS) version $13.5.0$ for the data analysis.

The  calibrated and concatenated event lists was generated using SAS task 
\textit{emproc} for EPIC MOS1 instruments. The events file was generated then subjected to 
filtering for background flaring using the SAS task \textit{evselect} and 
\textit{tabgtigen} identifying the intervals of background flaring and thus 
creating the Good Time Intervals (GTI) file. In this step, we applied the 
filtering expression in such a way that it selects the data with count rates of 
 $\leq 0.35$ counts s$^{-1}$  in energy range $> 10$ keV. The 
above correction to the event lists results the sum of all GTI  =$28.9$ for MOS1 camera.

\begin{figure}[h]
\begin{center}
 \includegraphics[scale=0.45]{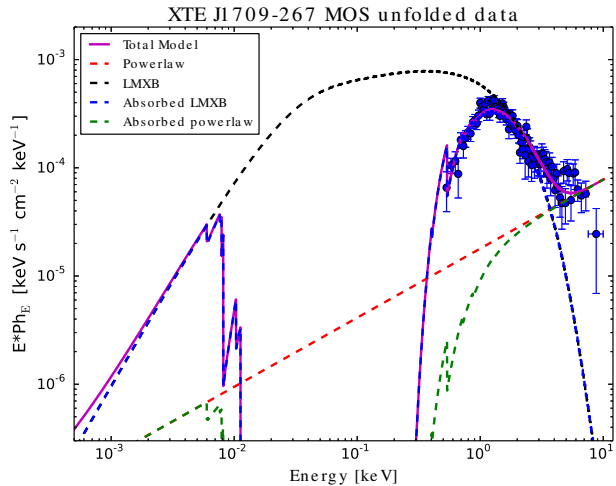}
\end{center}
 \caption{{\it XMM-Newton} MOS1 data of the source XTE J1709-267. Blue points represent the data, 
while total models is shown by magenta solid line. 
 Different model components are presented by dashed lines: blue and black -- absorbed and 
unabsorbed {\it lmxb}, green and red -- absorbed and unabsorbed {\it powerlaw} respectively.}
 \label{fig:fit}
 \end{figure}

The source spectrum was extracted using the SAS task \textit{evselect} 
from a circular region of 30 arcsec
radius with the center at right ascension and declination of the source. 
The background was extracted using the annular region 
between the radii 33 arcsec and 40 arcsec. The corresponding response files and 
ancillary files were generated using the SAS tasks \textit{rmfgen} and 
\textit{arfgen}. Finally, the spectral data were binned to contain 
minimum 20 counts per bin. The data looked exactly the same as in 
Degenaar et al. (2013).

We used {\sc xspec} fitting package, version 12.9.0\footnote{https://heasarc.gsfc.nasa.gov/xanadu/xspec/} for further data analysis. 
Our model, normalized by the distance of $D=10$~Mpc, was prepared as a table model in the 
FITS format, and 
for the purpose of this paper we named it {\it lmxb}. 

Results of our fitting procedure are listed in Table~1, and are shown in 
Fig.~4.  The quality of our fit is high since the reduced $\chi^2=1.32$, however
 for some of the fitted parameters {\sc xspec} generated indefinite errors. These errors are 
indicated by dashes in Table~1. The total fitted model is a sum of 
disk model attenuating neutron star {\it lmxb} plus {\it powerlaw} emission of low intensity. 
Galactic absorption is included via {\it tbabs} (Wilms et al. 2000). We obtained the best effective
 temperature of the neutron star $T_{\rm eff,NS}= 6.335 \times 10^6$ K, the disk accretion 
rate $ \dot m = 0.0104\, \dot m_{\rm Edd}$,
and the viewing angle $i = 19.4^\circ$. Nevertheless, due to strong absorption, and limited 
observational coverage, our model cannot be fully tested. We used additional {\it powerlaw} 
with the photon index $\Gamma = 0.36$ to explain hard X-ray tail. This fact indicates that the
 inner emitting source is hotter and we are out of parameter space with {\it lmxb} model. 
Furthermore, we have extracted also four data points from Optical Monitor (OM) on the board 
of {\it XMM-Newton}, but they were three orders of magnitude above our model. This means that 
optical signal is not associated with the emission from an outer disk. 

\begin{table}
\begin{center}
\caption{Parameters from the fitting of XTEJ1709-267 {\it XMM-Newton} MOS1 data. The total 
model is {\it tbabs*(lmxb+powerlaw)}. The reduced $\chi^2$ of the fit is equal 1.32. See the 
text for discussion.}
\label{tab:obs}
\begin{tabular}{llll}
\hline 
Model & Parameter &  Value & Error \\  \hline \hline
{\it tbabs}  & $N_{\rm H}$ &  $2.67 \times 10^{21}$ cm$^{-2}$       & $\pm   0.22$ \\ 
{\it lmxb} & $T_{\rm eff,NS}$ & $6.335 \times 10^6$ K   & $ \pm 0.105 \times  10^6$\\
{\it lmxb} & $\dot m $  & 0.0104 $\dot m_{\rm Edd} $   & $\pm 0.0019$ \\
{\it lmxb}  & $\theta$  &  $70.6^{\circ}$ & --\\ 
{\it lmxb}  & Norm.  &  $5.33 \times 10^{-5}$   & $\pm0.27 \times 10^{-5}$ \\ 
{\it powerlaw}  & $\Gamma$  &  0.36   & --\\ 
{\it powerlaw}  & Norm.  &  $1.79 \times 10^{-5}$   & -- \\ 
\hline 
\end{tabular}
\end{center}
\end{table}

Our fit is consistent with the result by Degenaar et al. (2013), where the X-ray data are 
attributed to emission from the neutron star atmosphere. According to our fit, the neutron 
star atmosphere is  3.5 times hotter, but the warm absorption column is by
0.2 smaller, than the same values reported by Degenaar et al. (2013). Unfortunately, the source 
chosen by us has no data coverage below 0.5 keV, where we expect the transition between two 
visible bumps. 
In addition, this part of spectrum is strongly modified by warm absorption. Nevertheless, 
we decided to present Fig.~4, when it is clear what we should do to verify our model.
 We have to build
the model with hot ($\sim 10^{8-9}$~K) compact corona as an inner emitting source. This will 
moves the transition between two bumps towards higher energies, and with good data coverage 
the test of our model will be possible.

\section{Conclusions}
\label{sec:concl}

In this paper we theoretically explained the appearance of non-spherical emission by proper 
computation of the amount of energy which goes directly towards observer.
We defined full set of terms for observed intensity of a non-spherical system, consisting of 
a neutron star and the accretion disk. We took into account attenuation effects as seen by a 
distant observer at various aspect angles. We demonstrated that the overall continuum shape 
shows two peaks. The lower energy peak is caused by the accretion disk emission, whereas higher 
energy bump is due to the neutron star.
The position of the transition between peaks and their relative normalization is self 
consistently computed by our formulae. Contrary to the spherically symmetric emission the 
neutron star contribution in our model depends on the aspect angle, due to attenuation effect.

  In this paper we showed, that if the observed intensity of a
source was correctly computed, even without relativistic correction
and without light from the boundary layer, then we obtain the double
bump source spectrum due to  the double non-spherically symmetric, emitting region. 
Such double bump spectrum, or the broadness of the spectral disk component,  are very often
 observed in X-ray band of accreting objects.

We attempted to fit the X-ray spectrum of LMXB XTE~J1709-267, downloaded from 
{\it XMM-Newton} archive. Nevertheless, the spectral data are too narrow co cover and fit our
 model, because we have no points below 0.5 keV. The additional points in optical and UV domains 
are necessary to discriminate our model parameters. 
Unfortunately,  the double bump structure predicted by our model of LMXB with an accretion disk 
is difficult to observe since it falls into a photon energy range less then 0.5 keV. 

If the non-spherical emitting system contains hot compact corona (Fabian et al. 2015) instead of 
the neutron star, the double peak structure should be visible in X-ray domain. 
We argue that for some parameters the total spectrum from both components can explain soft X-ray 
excess in those objects. Our model fully predicts the broadness of the disk component observed 
in some sources (Kolehmainen et al. 2011, and
priv. com.).

\Acknow{This research was supported by  Polish National Science 
2013/11/B/ST9/04528,   
2015/17/B/ST9/03422, 2015/18/M/ST9/00541 
and by Ministry of Science and 
Higher Education grant W30/7.PR/2013. It has received funding
from the European Union Seventh Framework Program (FP7/2007-2013) under 
grant agreement No.312789.}

\end{document}